\documentclass[a4paper]{IEEEtran}
\IEEEoverridecommandlockouts
\usepackage{cite}
\usepackage{amsmath,amssymb,amsfonts}
\usepackage{algorithmic}
\usepackage{graphicx}
\usepackage{textcomp}

\usepackage{amsmath}
\usepackage{amsfonts}
\usepackage{amssymb}
\usepackage{amsthm,amsmath}
\usepackage{mathrsfs}
\usepackage{indentfirst}
\ifCLASSOPTIONcompsoc
 \usepackage[caption=false,font=normalsize,labelfont=sf,textfont=sf]{subfig}
\else
 \usepackage[caption=false,font=footnotesize]{subfig}
\fi

\usepackage{xcolor}
\def\BibTeX{{\rm B\kern-.05em{\sc i\kern-.025em b}\kern-.08em
    T\kern-.1667em\lower.7ex\hbox{E}\kern-.125emX}}
\begin{document}

\title{Heterogeneous Seismic Waves Pattern Recognition \\in Oil Exploration with Spectrum Imaging}
\author{\IEEEauthorblockN{Wang Yuyang}
\IEEEauthorblockA{\textit{} \\
\textit{Beijing University of Technology,}
Beijing, China. \\
\textit{School of Computer Science,}
\textit{University College Dublin,}
Dublin, Ireland. \\
yuyang.wang821@gmail.com}

}
%

\maketitle

\begin{abstract}
The use of seismic waves to explore the subsurface underlying the ground is a widely used method in the oil industry, since different kinds of the rocks and mediums have different reflection rate of the seismic waves, so the amplitude of the reflected waves can unraveling the geological structure and lithologic character of a certain area under the ground, but the management and processing of seismic wave data often affects the efficiency of oil exploration and development. Different kinds of the bulk seismic data are always mixed and hard to be classified manually. This paper presents a classification model for four main types of seismic data, and proposes a classification method based on Mel-spectrum. An accuracy of $98.32\%$ was achieved using pre-trained ResNet34 with transfer learning method. The accuracy is further improved compared with the Fourier transform method widely used in previous studies. Meanwhile, the transfer learning method and fine-tune strategy to train the neural network by training the first $N-1$ layers of the network separately and then train the fully connected layers further improves the training efficiency. Our model can also be seen as an efficient data quality control scheme for oil exploration and development. Meanwhile, our method is future-proofed ,for further improvement of the seismic data processing quality control system, according to the spectrum characteristics, this model can be further extended into a problematic data classification model, which can identify more types of data problems, thus reducing the workload of the bulk data management.
\end{abstract}


\IEEEpeerreviewmaketitle

\section{Introduction}
\IEEEPARstart{T}{he} use of seismic waves to obtain strati-graphic structure is an effective and mature method, which has been widely used in the fields of oil exploration and urban planning\cite{8242387}. However, in the process of long-term business activities\cite{9664568}, a large amount of exploration results data has been accumulated. However the lack of a comprehensive process for standardized management and application in the early stage often causes bottlenecks in the organization, quality control, data storage management and analysis of results data\cite{7943496}.

The first applicable seismic pattern recognition research was purposed by Liu(1982) using the method of nearest-neighbor decision rule for syntactic patterns. However, because of the insufficient computational resource, this model are very sensitive to the features selected, the selection of training samples, and the weight assignment. The idea of using Neural network to solve the problems was purposed by Harrigan(1991) at the first time, a method based on Multi Layer Perceptron (MLP) were purposed for recognition\cite{140293}. Recently, several method based on convolutional neural network (CNN) and ensemble empirical mode decomposition (EEMD) were purposed, Li(2020) transformed the raw singal to Hilbert
spectrum for classification\cite{9353037}. Peng(2021) further extended the method using deep convolutional neural network (DCNN), instead of transform the singal to spectrum, they takes the raw singal as the input of the model\cite{9447036}.

\begin{figure}[!t]
	\centering
		\includegraphics[width=3.3in]{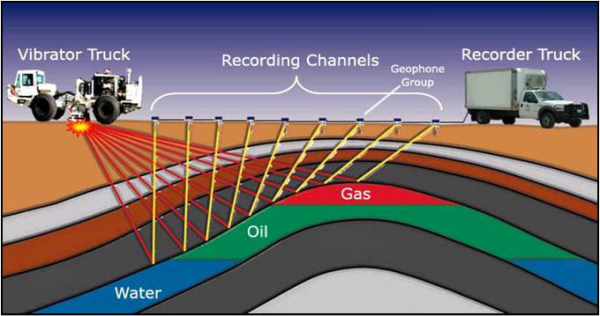}
		\caption{Seismic Exploration}
		\label{SeismicExploration}
\end{figure}

The previous works above promoted the seismic wave field separation, denoising and recognition obviously. In this paper, Seismic waves data files (e.g. SEG-Y Data Exchange Format) were converted into sets of Mel-spectrum. Our model has a relatively deeper layers and shot-cutting structure to let the layers preserve enough information of the inputting images, thus achieved high accuracy and low training cost. Meanwhile, the fine-tune method was adopted during the training process by giving a small leaning rate to let the network adapt the features of our seismic waves spectrum, which saves the training time while ensuring the accuracy\cite{5648084}.




\section{Methodology}

\subsection{Signal Acquisition}
The basic technique of seismic exploration involves generating seismic waves and measuring the time it takes for them to travel from the source to a series of geophones on the ground\cite{4539259}. By measuring the time of reaching each set of geophones as well as the amplitude of the reflected waves \cite{8308661}, after filtering and stacking, we can reconstruct the subsurface of the earth consistent with geological elastic discontinuities, which may indicate the presence of oil and gas Fig.\ref{SeismicExploration} illustrated how the seismic waves got recorded\cite{4300423}.

\begin{figure*}[!t]
\centering
\subfloat[Type I: Raw Migration Data]{\includegraphics[width=1.7in]{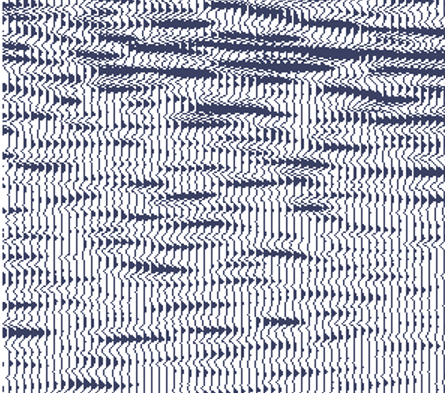}%
\label{fig_first_case}}
\hfil
\subfloat[Type II: Final Migration Data]{\includegraphics[width=1.75in]{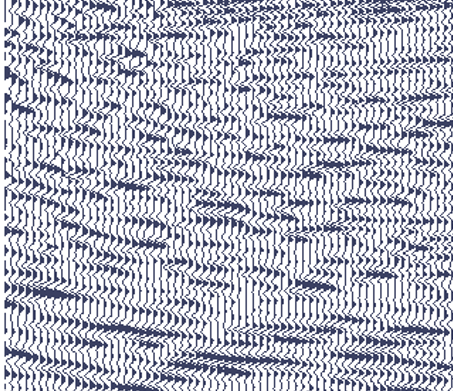}%
\label{fig_second_case}}
\hfil
\subfloat[Type III: Raw Stack Data]{\includegraphics[width=1.66in]{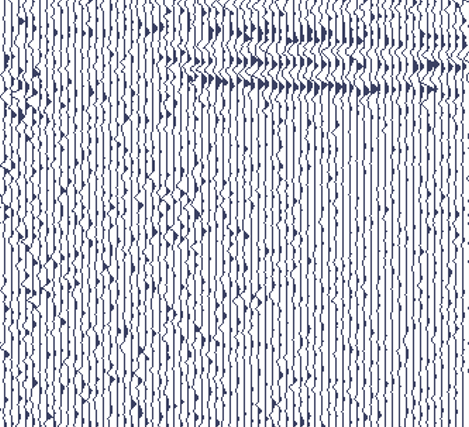}%
\label{fig_second_case}}
\hfil
\subfloat[Type IV: Final Stack Data]{\includegraphics[width=1.68in]{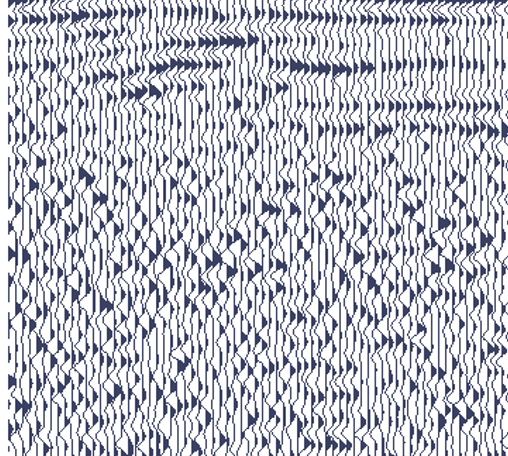}%
\label{fig_second_case}}
\caption{Four types of seismic waves.}
\label{four_types}
\end{figure*}

\begin{figure}[!t]
	\centering
		\includegraphics[width=3.0in]{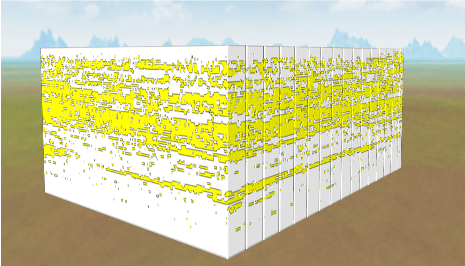}
		\caption{3D stratigraphic structure reconstruction}
		\label{GroundStructure}
\end{figure}

\begin{figure}[!t]
	\centering
		\includegraphics[width=3.0in]{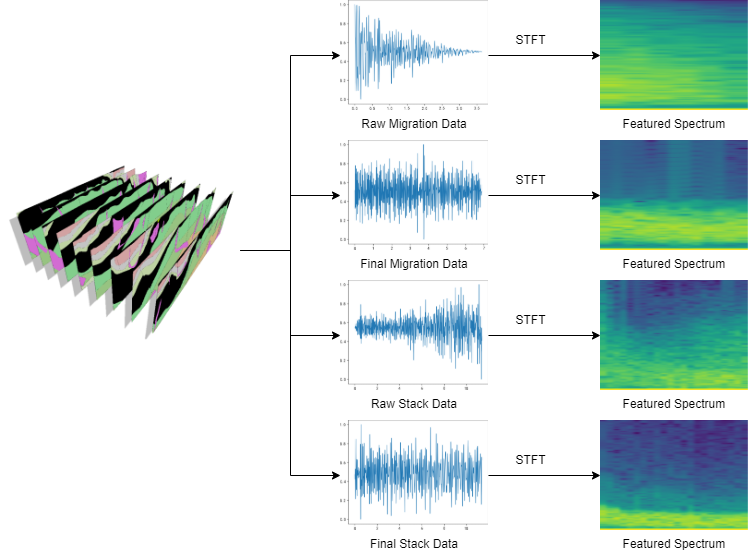}
		\caption{Waveform-Spectrum Transformation}
		\label{Waveform-Spectrum}
\end{figure}

The signal received by the geophones can be used to infer the composition of subsurface structure, shown in Fig.\ref{GroundStructure}. Before that, signal pre-processing is required , including filtering, stacking and amplification\cite{1315021}. According to this, we can classify the seismic wave data into four categories.

\begin{itemize}
    \item Raw Migration Data
    \item Final Migration Data
    \item Raw Stack Data
    \item Final Stack Data
\end{itemize}

These four wave-forms shows in \textit{Fig.\ref{four_types}} are often mixed in seismic data management and processing\cite{4740813}.

\subsubsection{Raw data}
Before stacking, the energy intensity of the seismic effective wave signal is maintained with amplitude fidelity, but the data are still mixed with noise, which shows stronger energy in the high-frequency part on the spectrum, and the data shows an unsmoothed waveform curve with more local spikes\cite{6288428}. The homophase axis continuity is poor. The shallow energy is strong and the deep energy is weak.

\subsubsection{Stack data}
The processing of seismic data acquired by multiple coverage method, the records of many channels with common reflection points are stacked after dynamic correction to improve the signal-to-noise ratio and suppress the interference, and the seismic profile obtained by this method is called horizontal stacked profile, or stacked profile for short.

\subsubsection{Migration data}
After horizontal stacking, the reflection layer is automatically spatially homed, and the seismic profile obtained by this method is called stacked migration data, or migration data for short\cite{5555664}.

\subsubsection{Final data}
Final data are generally modified (filtered and dynamically balanced), the noise is weaker, and the spectrum shows weaker energy in the low and high frequency parts outside the effective frequency band. The homophase axis continuity is significantly increased in this kind of profile, the deep and shallow energy are further balanced. 

In the management of seismic data, especially the management of massive legacy data, it is often the case that the data type labels are wrong, missing or inconsistent with the actual data content, in the field of seismic data quality control, it is necessary to have massive inspection on seismic data consistency and continuity, which is time-consuming and labor-intensive for manually checking. To find a method which overcomes the challenges mentioned above has signification meaning and value for research and practical application.

\subsection{Data Pre-Processing}
Considering that the amplitude of the original acquired original seismic wave signal is very weak and mixed with a relatively large amount of noise\cite{6384291}, which significantly affects the process of feature extraction, thus pre-processing of the original signal is required. The range of seismic wave frequencies are between $2-90Hz$, where the frequency of $20-70Hz$ contains the most valuable data\cite{8755388}, so first we filtered the data to eliminate the low frequency noise and high frequency noise. The high-pass and low-pass filter can be represented by Eq.\ref{loss-pass} and Eq.\ref{high-pass}.

\begin{equation}
\label{loss-pass}
H_{L P}(j \omega)=\frac{1}{\tau j \omega+1}
\end{equation}

\begin{equation}
\label{high-pass}
H_{H P}(j \omega)=\frac{1}{\tau \frac{1}{j \omega}+1}
\end{equation}

To further enhance the data features, we remapped the filtered data to the range of $INT16$ and re-sampled the data in order to match the sample rate of seismic waves detector, thus making the features significant in the spectrum\cite{xuejun1}. After that, the data can be further augmented and classified.

\subsection{Feature Extraction and Augmentation}
Fast Fourier Transform (FFT) is a method to convert the time-domain map of a finite and stable signal into a frequency-domain map\cite{599273}. However, FFT cannot analyze the frequency domain features of infinite and unstable signals with time variation, such as seismic waves signals. The Short-Time Fourier Transform (STFT) is used to analyze how the frequency content of a nonstationary signal changes over time\cite{4814507}. Which can be used to display both time-domain features and frequency-domain features when time-frequency features are needed.

The STFT of a signal is calculated by sliding an analysis window of length $M$ over the signal and calculating the discrete Fourier transform of the windowed data. The number of columns in the STFT matrix is given by Eq.\ref{eq1}.
\begin{equation}
    \label{eq1}
    k=\left\lfloor\frac{N_{x}-L}{M-L}\right\rfloor
\end{equation}
The STFT matrix is given by 
\begin{equation}
    \mathbf{X}(f)=[
    \begin{array}{lllll}
       X_{1}(f) & X_{2}(f) & X_{3}(f) & \cdots & X_{k}(f)
    \end{array}]
\end{equation}
such that the $m_{th}$ element of this matrix is
\begin{equation}
    \label{eq2}
    X_{m}(f)=\sum_{n=-\infty}^{\infty} x(n) g(n-m R) e^{-j 2 \pi f n}
\end{equation}
where
\begin{itemize}
    \item $g(n)$: Window function of length $M$
    \item $X_m(f)$: DFT of windowed data centered about time $mR$
    \item $R$: Hop size between successive DFTs.
\end{itemize}
 
To perform the STFT transformation we choose TorchAudio, which is included in PyTorch as a library, the transformation details has been demonstrated in Fig.\ref{Waveform-Spectrum}. The next stage of our model is to train the network with the spectrum image as the input, it is necessary to make the input of the network be a fixed size, so the size of the spectral image output by the STFT algorithm was fixed to $256 \times 256$. 

\begin{figure}[!t]
	\centering
		\includegraphics[width=2.5in]{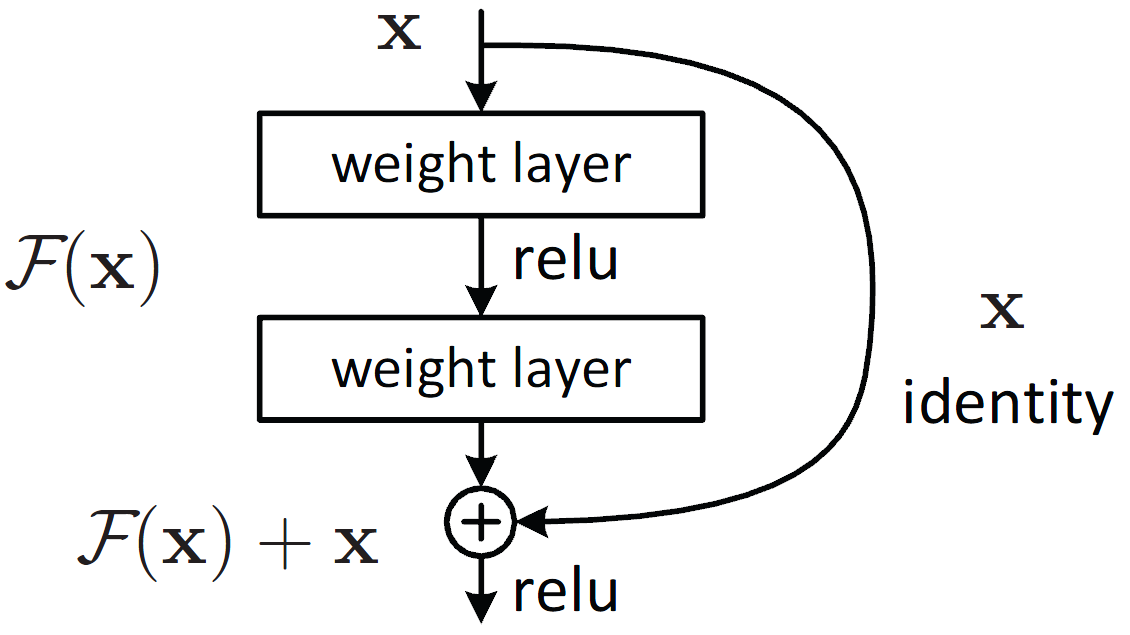}
		\caption{Short-cutting in ResNet}
		\label{Short-cutting}
\end{figure}
\begin{figure}[!t]
	\centering
		\includegraphics[width=2.3in]{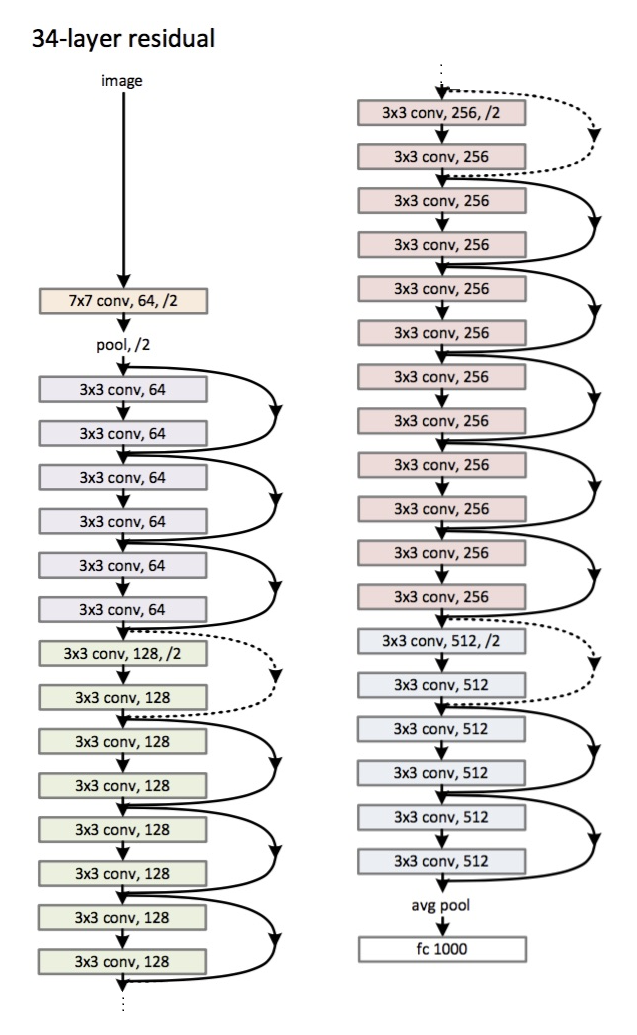}
		\caption{ResNet34 Baseline}
		\label{ResNet34_baseline}
\end{figure}

\section{Feature Classification}
After obtaining the spectrum images of the four types of seismic wave data, we brought these four types of images into deep residual network for training. The deep residual network (ResNet) solves the phenomenon that the accuracy of the training set decreases as the network deepens\cite{he2015deep}. ResNet proposes a neural network structure similar to short-circuiting and incorporates residual units through a short-circuiting mechanism shown in Fig.\ref{Short-cutting}, so the degradation problem is well solved. According to the advantages of the ResNet, we chooses this network as the baseline of our model shown in Fig.\ref{ResNet34_baseline}.

In the problem we are solving, there are four types of spectrum that need to be classified, so we need to replace the number of neurons in the fully connected layer so that the network can be adapted to our training data. Here we replace the original 1000 classes of the fully connected layer with 4 classes. To further reduce the training time, we choose transfer learning approach to improve the training efficiency by using a pre-trained neural network and choosing a relatively low learning rate to allow the model to fully absorb the features of the new data. In the actual training process, we freeze the results of the first $n-1$ layers and train the fully connected layers separately, and then freeze the fully connected layers to train the first $n-1$ layers separately after reaching a certain accuracy, thus achieving a higher accuracy rate with the same training time.

\section{Result and Discussion}

\begin{figure*}[!t]
\centering
\subfloat[Training accuracy curve]{\includegraphics[width=2.95in]{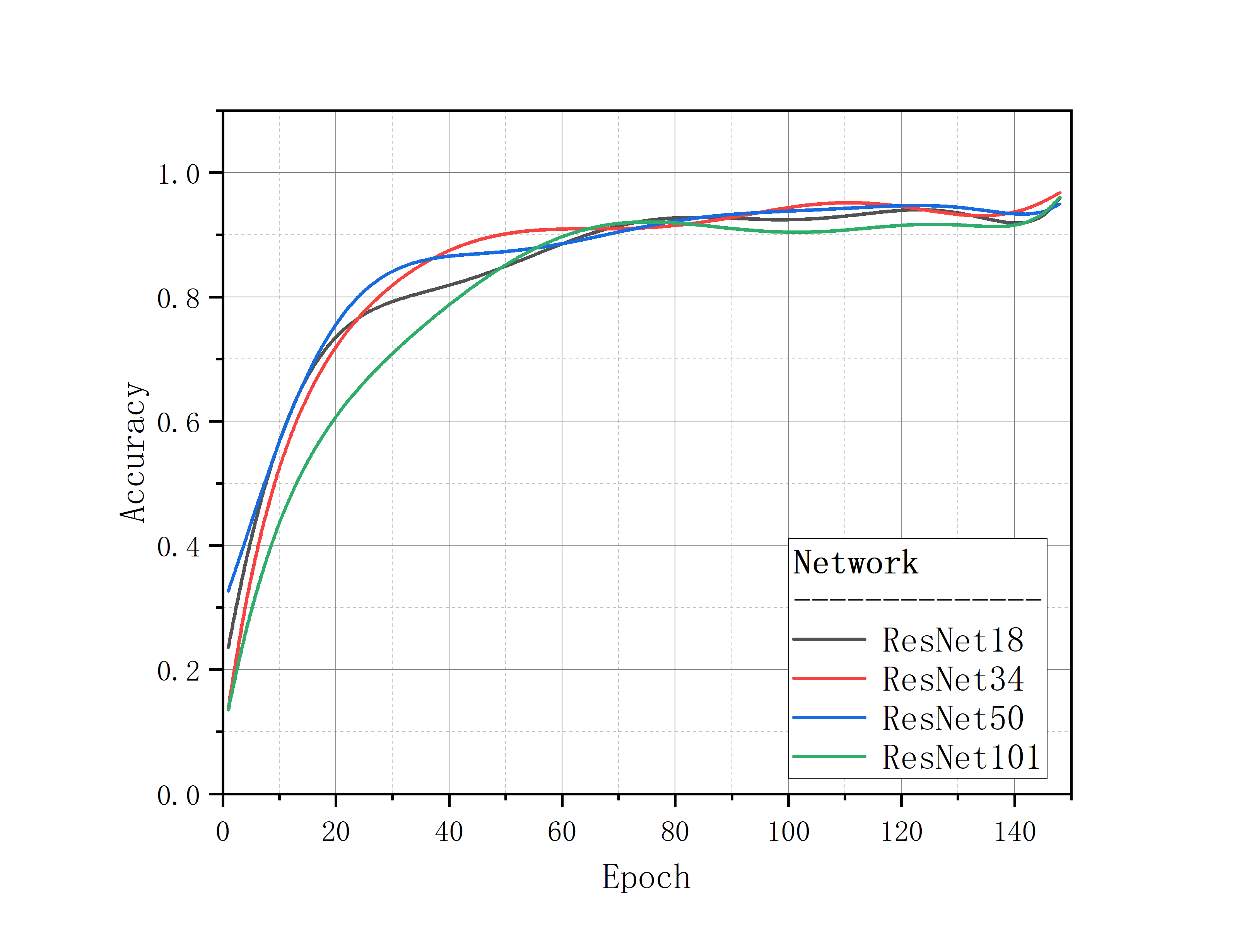}%
\label{fig_first_case}}
\hfil
\subfloat[Training loss curve]{\includegraphics[width=2.95in]{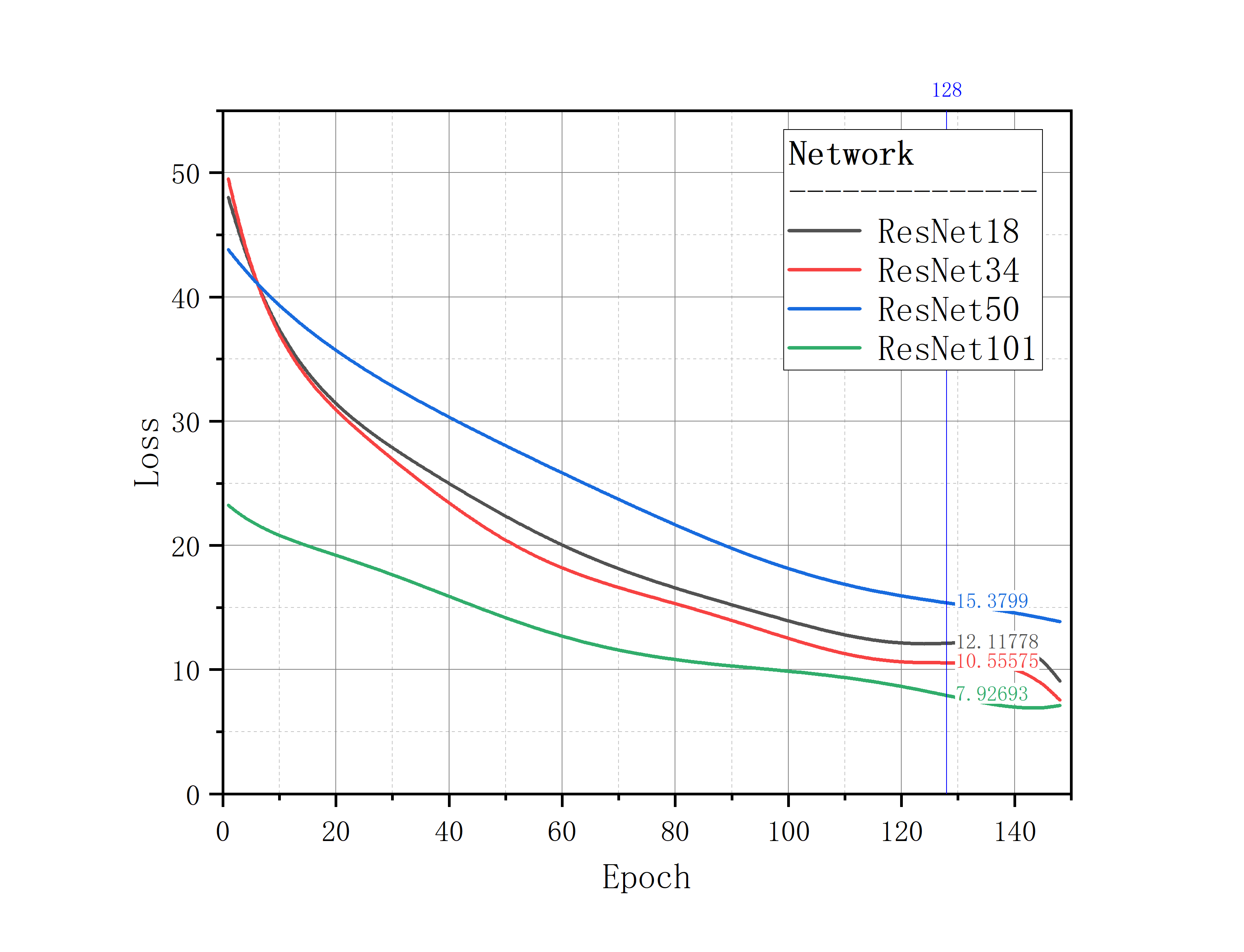}%
\label{fig_second_case}}
\caption{Training results for the network.}
\label{TrainingdifferentResNet}
\end{figure*}

\begin{figure}[!t]
	\centering
		\includegraphics[width=2.9in]{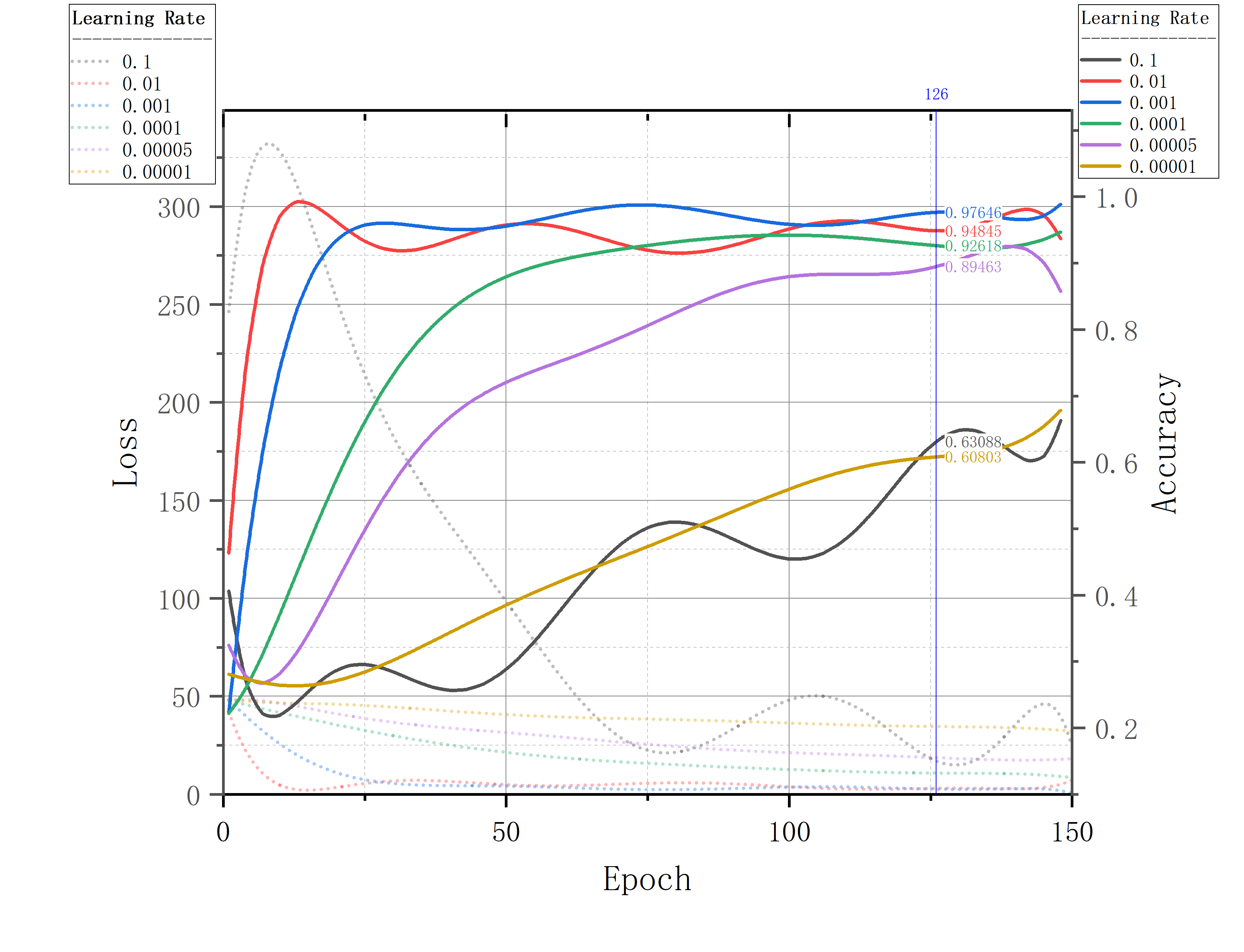}
		\caption{ResNet34 Epoch-accuracy curve of different Learning Rate}
		\label{Epoch-accuracyLearningRate}
\end{figure}

Since the training time and computation required of different depths of neural networks vary, different training on ResNet18, ResNet34, ResNet50 and ResNet101 were conducted to compare the performance of different depths of neural networks on the seismic data-set in order to to select the best neural network model.


\begin{table}[!t]
\renewcommand{\arraystretch}{1.3}
\caption{Comparison of Different Network Model}
\label{ComparisonDifferentModel}
\centering
\begin{tabular}{lllll}
\hline
Network model & ResNet18 & ResNet34 & ResNet50 & ResNet101 \\ \hline
Accuracy      & $96.48\%$  & $98.22\%$  & $97.38\%$  & $98.18\%$   \\
Average loss  & $9.19$     & $8.39$     & $17.46$    & $8.32$      \\
Learning rate & $0.001$    & $0.001$    & $0.001$    & $0.001$     \\ \hline
\end{tabular}%
\end{table}
\begin{table}[!t]
\renewcommand{\arraystretch}{1.3}
\caption{Training Result of ResNet34 at different Learning Rate}
\label{ResNet34LearningRate}
\centering
\begin{tabular}{lll}
\hline
Learing Rate & Accuracy & Average loss \\ \hline
$0.1$          & $68.43\%$  & $34.84$       \\
$0.01$         & $94.23\%$  & $8.93$         \\
$0.001$        & $98.22\%$  & $8.39$         \\
$0.0001$       & $96.17\%$  & $10.38$        \\
$0.00001$      & $69.93\%$  & $25.32$        \\ \hline
\end{tabular}%
\end{table}

The results in Fig.\ref{TrainingdifferentResNet} and Table.\ref{ComparisonDifferentModel} illustrates a comparison of models with different number of layers, ResNet18 still has a higher loss after $120$ Epochs due to its shallow depth, ResNet101 has the deepest depth and therefore has a slower increasing slop in accuracy comparing to the other networks at the beginning of training, and it's loss is the lowest among all the models. However, due to it's relatively large number of parameters, it took longer time to converge in the training process. Finally, we chose ResNet34 as the best neural network, which has a moderate number of total parameters and a fast convergence time and high inference accuracy. It even outperformed ResNet50 at $140$ Epoch in terms of loss.

\begin{table}[!t]
\renewcommand{\arraystretch}{1.3}
\caption{Classification Accuracy of Four Types of Wave}
\label{ComparisonWaves}
\centering
\begin{tabular}{lllll}
\hline
Wave Type     & Type I & Type II & Type III & Type IV \\ \hline
Type I       & $1.0$     & $0.95$    & $0.99$     & $0.97$   \\ 
Type II      & $0.95$    & $1.0$     & $0.98$     & $0.97$   \\ 
Type III     & $0.99$    & $0.98$    & $1.0$      & $0.99$   \\ 
Type IV      & $0.97$    & $0.97$    & $0.99$     & $1.0$   \\ \hline
\end{tabular}%
\end{table}

Meanwhile, we compared the convergence speed and prediction accuracy of ResNet34 at different learning rates shown in Fig.\ref{Epoch-accuracyLearningRate} and Table.\ref{ResNet34LearningRate}. It can be seen that at learning rates of $0.0001$ and $0.1$, the convergence speed is slow because the learning rate is too low and the network cannot find the local minimum quickly, while the latter has a large fluctuation in network accuracy because the learning rate is too high and cannot stay near the minimum, so the final accuracy is not high either. Finally, we choose $0.001$ as the training learning rate of the network, which has a stable accuracy increase rate and a faster convergence rate while the accuracy after convergence can also meet the requirements. We also compared the model's capability of differentiating 4 types of waves, result has been shown in Table.\ref{ComparisonWaves}. From the result we can see that our model can perfectly identify most of the features, but sometime it might fail e.g. Type.I and Type.II both have similar features at peak data points.


\section{Conclusion}
To summarize, the method we proposed is able to classify the four types of seismic wave signals with high accuracy. We demonstrate that the classification task of seismic wave signals can be achieved using spectrum analysis as well as the use of high-pass filter and low-pass filter. The use of a lower learning rate to achieve transfer learning also reduces the training time and facilitates faster convergence of the model when the amount of data is larger, thus enhancing the robustness of the model. In the future, the spectrum data can be further processed, including more detailed filtering and feature enhancement, so that more types of seismic wave information can be identified.

\bibliographystyle{IEEEtran}
\bibliography{bibtex/mybib}

\end{document}